\documentclass[11pt]{article}
\usepackage{epic}

\usepackage{amsmath,amssymb}
\usepackage{amsthm}
\usepackage{amsfonts}
\usepackage{amscd}
\usepackage{mathrsfs}
\usepackage{eufrak}

\newcommand{\be}{\begin{equation}}
\newcommand{\ee}{\end{equation}}
\newcommand{\ba}{\begin{eqnarray}}
\newcommand{\ea}{\end{eqnarray}}

\def \g{\gamma}

\def\P{{\cal P}}

\title{QCD-like properties for anomalous dimensions in $\mathcal{N}=4$ SYM}

\author{Valentina Forini $^{a}$ and Matteo Beccaria $^{b}$ \\
$^{a}$ Max-Planck-Institut f\"ur Gravitationsphysik, \\
    Albert-Einstein-Institut,  Am M\"uhlenberg 1,  \\
    D-14476 Potsdam, Germany\\
$^{b}$Physics Department, Salento University and INFN, 73100 Lecce, Italy \\
E-mail: forini@aei.mpg.de, beccaria@le.infn.it}

\begin{document}
\maketitle

\begin{center}
\begin{minipage}[h]{10cm}
\centerline{\bf Abstract}
\medskip
In this contribution it is shown how closed formulas for anomalous dimensions of two classes of operators in $\mathcal{N}=4$ SYM can be derived, either by investigating the numerics or on the basis of QCD-inspired assumptions. We discuss  the case of twist three ``gauge'' operators in which  a complete proof of reciprocity can be carried out.
\end{minipage}
\end{center}



\section{Introduction and conclusions}

Maximally supersymmetric theory ${\cal N}=4$ SYM is dual to type IIB superstring on $AdS_5\times S^5$ and plays a central role in the AdS/CFT correspondence~\cite{Maldacena:1997re}. 
Recent developments in the study of the duality between the planar, large N limit of the gauge theory and the free string theory are based on the development of analytic tools that exploit the integrability of the string side~\cite{Bena:2003wd} as well as an {\em internal} integrability of the superconformal theory~\cite{Beisert:2004ry}. In the latter case, the scale dependence of renormalized composite operators is governed, even at higher loops, by a local, integrable, super spin chain Hamiltonian whose interaction range increases with the loop order~\cite{Beisert:2003yb,Beisert:2005fw}. This fact has set  the long range asymptotic Bethe equations of Ref.~\cite{Beisert:2005fw} as a natural tool for calculating anomalous dimensions of single traces operators of the general form 
\be\label{operator}
{\cal O} = \mbox{Tr}\,\left(\,\prod_{i=1}^L D^{n_i}\, X_i\right) + \mbox{permutations},
\ee
where $X_i$ are elementary fields in certain subsectors of the full ${\cal N}=4$ SYM and $D^{n}$ are covariant derivatives. 
The Bethe equations provide the anomalous dimensions of  ${\cal O}$ as a perturbative series in the 't Hooft coupling $g$
\be\label{anomalous}
\g_{\cal O}(g)=\sum_{n\geq0}\,c_n({\cal O})\,g^{2n}
\ee
and stand as a formidable tool for multi-loop calculations otherwise drastically hard to face. 
However, their asymptotic nature   leads to the major limitation of ``wrapping'', for which $\g_{\cal O}$ is actually calculable up to ${\cal O}(g^{2L})$ terms (being $L$ the length of the operator). While in the thermodynamical limit, in which the  $L $ or the Lorentz spin $N =\sum_i\{n_i\}$ are sent to infinity,  the wrapping problem can be overtaken, a more subtle limitation appears when one tries to investigate the dependence on the parameters above at fixed perturbative order. In general, the Bethe equations do not provide the expansion coefficients $c_n$ in (\ref{anomalous}) as functions of  $L$ and $\{n_i\}$, but furnish just sequences of numerical (sometimes rational) values for each given operator. In order to find \emph{closed formulas}  for the anomalous dimensions of interest thus opening the way to a deep investigation of their properties,  one has to resort either to conjectures that need further numerical confirmation or, as in the case of $\g(N)$,  to some general (typically QCD-inspired) assumption.

A first example is the study of Ref.~\cite{Beccaria:2007gu}, in which operators of the type ${\cal O}_L={\rm Tr}{\cal F}^L$, built of $L$ components of the self-dual Yang-Mills field strength, have been considered~\footnote{These operators are exact eigenstates of the one-loop dilatation operator and can be mapped to the ferromagnetic states of an integrable spin $S=1$ chain~\cite{Ferretti:2004ba,Beisert:2004fv}. At two loop and beyond, they mix with the other $\mathfrak{psu}(2,2|4)$ fields.}.
Beyond one-loop, the analysis of Ref.~\cite{Rej:2007vm} provide efficient computational tools to derive the sequence $\{c_n(L)\}$ for any given $L$, although not parametrically. It turns out immediately that $c_n(L)$ is not linear in $L$ as far as $n\ge 2$. So, starting at two-loops, non-trivial finite size corrections appear and the ratio $\g(L)/L$ is not expected to be a simple expression at finite $L$. However, some unexpected structure exists. A careful investigation of the (infinite precision) numerics for the sequences $\{c_n(L)\}$ at fixed $n$ and $L$ varying  has been crucial for conjecturing and providing closed expressions, up to 5 loops, for $c_n(L)$ in (\ref{anomalous}).  Such a closed formula for the multi-loop \emph{size dependence} have no counterpart in existing calculations for other operators in the various subsectors of ${\cal N}=4$ SYM.
To give an example, the two-loop anomalous dimension takes the remarkably simple form 
\be
\gamma_L(g) = 2\,L+3\,L\,g^2 + \left(-\frac{51}{8}+\frac{9}{8}\,\frac{1}{(-1)^L\,2^{L-1}+1}\right)\,L\,g^4 + \cdots\, ,
\ee
with exponentially suppressed corrections to the trivial linear scaling with $L$. Remarkably, its large $L$ limit
\be
\frac{\gamma_L(g)}{L} = f_0(g) + g^4\, h(g\,L)\, e^{-L\,\log\,2} + {\cal O}(e^{-2\,L\,\log\,2})
\ee
shows that the size corrections to the thermodynamical limit are characterized by a finite $g$-independent correlation length $\xi = 1/\log\,2$ and the combination $g\,L$ as
a natural scaling variable for the prefactor. Such ${\cal O}(2^{-L})$ corrections have nothing to do with much smaller ${\cal O}(\lambda^L)$ wrapping effects, and it would be interesting to understand them from the point of view of the spin-chain interpretation of the dilatation operator $H$~\footnote{In particular, a natural explanation for the exponential corrections could take into account length-changing processes, as suggested in Ref.~\cite{Rej:2007vm}, and an explicit two-loop calculation of $H$ would be important to clarify these issues.}.

In the following, we will focus on a second particularly interesting class of operators, the  so-called \emph{quasipartonic} twist operators~\cite{quasipart}. These are single trace operators of the type (\ref{operator}) constructed with an arbitrary number of  derivatives (in this case projected on the light-cone) distributed among (collinear) twist $1$ fundamental fields $X$ (scalars, gauginos or gauge fields), in such a way that the length $L$ of the operator coincides with the twist of the operator. Quasipartonic twist operators are  interesting because of the similarities with  twist operators in QCD. Indeed, while ${\cal N}=4$ SYM and QCD are in many details different, a compared analysis of  their properties has been crucial for a deeper understanding of  both of them.
Integrability itself, as a basis for the evolution of composite operators, has been first discovered in the study of planar QCD~\cite{QCD-int}. About conformal symmetry, unbroken in QCD at one loop, it does  not appear to be a necessary condition for  integrability, as discussed in Ref.~\cite{Belitsky:2004sc,Belitsky:2004sf,Belitsky:2005bu,DiVecchia:2004jw}, but it certainly plays an important role by imposing selection rules and multiplet structures. Moreover, it is a (somewhat hidden) consequence of conformal symmetry that can explain the structure of the large spin expansion of the twist operators anomalous dimensions (see footnote ${\rm h}$). It is also believed that QCD would benefit a lot from an ultimate all-loop solution of its superconformal version, since this would provide a representation for the ``dominant'' part of the perturbative gluon dynamics (see, for example, Ref.~\cite{Dokshitzer:2008zz})~\footnote{Other notable common issues between ${\cal N}=4$ SYM  and QCD , such as their infrared structure, are reviewed in Ref.~\cite{Dixon:2008tu}.}.

The  \emph{maximum transcendentality principle} is a first interesting example of such an interplay between the theories. Inspired by the structure of the two loop anomalous dimension of $\mathcal{N}=4$ twist two operators in the $sl(2)$ sector, Kotikov, Lipatov, Onishchenko and Velizhanin~\cite{Kotikov:2004er} have proposed that the three-loop answer could be extracted by simply picking up the ``most transcendental terms''  from the three-loop non-singlet QCD anomalous dimension derived f in Ref.~\cite{Moch}. The conjectured three loop formula has been then independently confirmed in the framework of the Bethe ansatz equations ~\cite{Staudacher:2004tk} as well as within a space-time approach~\cite{Bern:2005iz}.
The  principle, according to which \emph{$\gamma^{(n)}(N)$ at $n$ loops is a linear combination of Euler-Zagier harmonic sums of transcendentality $\tau=2n-1$}, has been the key via which closed multi-loop $N$-dependent expressions for the anomalous dimension of special twist operators have been then derived~\cite{Staudacher:2004tk,Kotikov:2007cy,Beccaria:2007cn,Beccaria:2007vh,Beccaria:2007bb,Beccaria:2007pb}. A systematic derivation of the functions $\gamma(N)$ beyond the one-loop level  can be done exploiting the  Baxter approach~\footnote{Talk given by S. Zieme at AEI, Potsdam, based  on a work in progress by A.~V.~Kotikov, A. Rej and  S. Zieme.}. Recent analytical attempts are discussed in Refs.~\cite{Belitsky:2007jp,Beccaria:2007uj}.

\emph{Closed expressions} of twist anomalous dimensions are crucial to investigate their physical content, that can be extracted by exploiting known facts valid for the QCD  twist (two) operators arising in the analysis of deep inelastic scattering~\cite{Altarelli:1981ax,Martin:2008cn}.
In that context, the total spin $N$ is dual, in Mellin space, to the Bjorken variable $x$ and two opposite regimes naturally emerge, small $x\to 0$ and $x\to 1$.
The first is captured by the BFKL equation~\cite{Lipatov:1976zz}, and  can be analyzed by considering 
the Regge poles of $\gamma(N)$ analytically continued to negative (unphysical) values of the spin. The BFKL equation has been the crucial testing device for detecting wrapping effects when Bethe equations are used to calculate anomalous dimensions of short operators ~\cite{Kotikov:2007cy}~\footnote{In the BFKL picture an interesting interpretation of the spin chain magnon has been recently given in Ref.~\cite{Gomez:2008hx}.}.

The properties of the second ({\em quasi-elastic}) regime,  which in Mellin space is  equivalent to the large $N$ limit, are the ones of interest here. They can be inferred from the large $N$ behavior of known three loop twist-2 QCD results (as well as from general results valid at higher twist~\cite{Belitsky:2003ys}) and can be summed up in two main points:
\begin{enumerate}
\item 
The \emph{leading} large $N$ behavior of anomalous dimensions for twist operators is logarithmic \be
\gamma(N) = 2\, \Gamma(\alpha_s)\,\log\,N + {\cal O}(N^0),\qquad N\to\infty,~~~
\ee
and governed by the the so called ``cusp anomaly'' $\Gamma(\alpha_s)$, a universal function of the coupling related to soft gluon emission~\cite{Korchemsky:1992xv, Belitsky:2003ys,Belitsky:2006en} and appearing as a cusp anomalous dimension governing the renormalization of a light-cone Wilson loop. 
Integrability techniques have drastically improved this knowledge providing an integral equation that furnishes the all-order weak coupling expansion~\footnote{The calculation has been extended at strong-coupling in the explicit case of the $\mathfrak{sl}(2)$ sector~\cite{Basso:2007wd} (see also Ref.~\cite{StrongSL2}) and is amenable to wide generalizations~\cite{frs, Kruczenski:2008bs}.} of $\Gamma(\alpha_s)$~\cite{Eden:2006rx,Beisert:2006ez}.
\smallskip
\item 
About \emph{subleading}  terms, it is  known that they obey (three loops) hidden relations, the Moch-Vermaseren-Vogt (MVV) constraints~\cite{Moch}. In the  twist-2 QCD context  such relations can be related with space-time reciprocity of deep inelastic scattering and its crossed version of $e^+e^-$ annihilation into hadrons. Technically, reciprocity in the twist-2 case  holds for the Dokshitzer-Marchesini-Salam (DMS) evolution kernel governing simultaneously the distribution and fragmentation functions~\cite{Dokshitzer:2005bf}~\footnote{The DSM  evolution kernel has recently received a nice confirmation in Ref.~\cite{Laenen:2008ux}.}. In Ref.~\cite{Basso:2006nk}, the MVV relations have been extended to an infinite set of  higher orders relations in the $1/N$ expansion, and their origin has been indicated to follow from the invariance under the $sl(2, \mathbb{R})$ subgroup~\footnote{Quasipartonic operators can be classified according to representations of the collinear $sl(2, \mathbb{R})$ subgroup of the
   $SO(2,4)$ conformal  group which are labeled by the so-called
    \emph{conformal spin}
     $j=\textstyle{\frac{1}{2}}(N+\Delta)$~\cite{quasipart}, where $\Delta$ is the scaling dimension of the operator. From this one may argue that the anomalous dimension $\g=\Delta-N-L$  should be a function
     of  the Lorentz spin $N$ only through its dependence on the conformal spin $j$.
     Since  $\Delta$ is  $\Delta = N + L+ \g(N,L)$  that then leads to a relation of the type (\ref{nonlinear}), where the function $\P$ depends on the twist $L$.}. 
\end{enumerate}

More specifically, a suitable generalization of the analysis of Refs.~\cite{Dokshitzer:2005bf,Basso:2006nk} to the case of $\mathcal{N}=4$ SYM  assumes that $\g(N)$ obeys at all orders the non-linear equation
\be
\label{nonlinear}
\gamma(N) = \P\left(N+\frac{1}{2}\gamma(N)\right),
\ee
where the function $\P$ has a large $N$ expansion in {\em integer powers of } $J^2$ of the form
\be
\label{parity}
{\cal \P}(N) = \sum_n\frac{a_n(\log\,J)}{J^{2n}},
\ee
Above, $J$  is the Casimir of the 
collinear conformal subgroup $SL(2, \mathbb{R})\subset SO(4,2)$, that is $
J^2 = (N+L\,s-1)\,(N+L\,s)$, 
where the   $s=1/2,1,3/2$ distinguishes between the scalar, spinor or vector case~\cite{quasipart}. If the expansion (\ref{parity}) holds, one says that ${\cal P}$
is a {\em reciprocity respecting}~\footnote{The name \emph{reciprocity}
comes
  from the formulation of this property for  the  Mellin  transform:
   \ $\tilde{P}(x)=-x\,\tilde{P}(1/x)$, where $\P(N)=\int_0^1dx\,x^{S-1}\,\tilde{P}(x)$.} (RR) kernel.
Beyond one loop, a test of reciprocity requires the knowledge of the multi-loop anomalous dimensions as closed functions of $N$. These are currently available in the cases of twist-2 and 3. Three-loop tests of reciprocity for QCD and for the universal
 twist 2 supermultiplet in $\mathcal{N}=4$ SYM were discussed
 in Refs.~\cite{Basso:2006nk,Dokshitzer:2005bf}. A four-loop test for
the   twist 3  anomalous dimension in the $sl(2)$ sector
was performed   in Ref.~\cite{Beccaria:2007bb}.  In Ref.~\cite{BF}  it was  proved that even the wrapping-affected
four loop result for the twist two operators ~\cite{Kotikov:2007cy} is reciprocity respecting in the sense of (\ref{parity}). This certainly suggests some important structure built in the Bethe Ansatz and deserving a deeper understanding. Indeed, while for  (\ref{nonlinear})  a relation to the underlying conformal symmetry has been suggested in Ref.~\cite{Basso:2006nk}, there is no obvious explanation for the property (\ref{parity}). 

Below, we illustrate the example of a four-loop  anomalous dimension obtained in closed formula by  exploiting  a (generalized) maximum transcendentality principle and successively analyzed in its structure to verify the RR relations. The  example of twist three gluonic operators is interesting for various reasons. First, at variance with the quasipartonic operators built with scalars and gauginos which belong to close sectors and thus scale autonomously at all loops~\footnote{Operators built of scalars
belong to the ${\cal N}=4$ $\mathfrak{sl}(2)$ subsector which is closed at all orders. About operators built of gauginos, they appear in the closed $\mathfrak{sl}(2|1)$
subsector where there is mixing between scalars and fermions, but not for the maximally fermionic component~\cite{Belitsky:2007zp}, which is the one of interest for the class of quasipartonic operators.},  the description as a {\em gluonic} operator is only correct at one-loop~\cite{Belitsky:2008wj} with mixing effects at 
higher orders (see the discussion in Ref.~\cite{Beccaria:2007pb}). Second, in the twist three case operators built with scalars, gauginos or gauge fields are not related by supersymmetry, at variance with the twist-2 case where all channels are in a single supermultiplet~\footnote{Remarkably, such twist two universality class is inherited in the gaugino sector~\cite{Beccaria:2007vh}.}. This richer multiplet structure has as a consequence the existence of various universality classes of anomalous dimensions, as well as the presence of a \emph{generalized} form of the maximum transcendentality principle.

As a final comment, we remark that it is natural to employ the AdS/CFT correspondence to investigate the presence of MVV-like relations at strong coupling. Since the planar perturbation theory should be convergent, such an organized structure of subleading terms in the large spin expansion should be visible also in the energies of the semiclassical string  states corresponding to twist operators. This analysis, initiated in Ref.~\cite{Basso:2006nk} for the folded string  at the classical level, has been recently extended in Ref.~\cite{BFTT} to configurations  (spiky strings)  that should correspond to twist operators with higher dimension and at one loop in string perturbation theory. Remarkably, the large spin expansion of the classical string energy happens to have exactly the same structure as that of $\g(N)$ in the perturbative gauge theory, and does respect MVV-like relation at one-loop. This provides a strong indication that these relations hold not only in weak coupling (gauge theory) but also in strong coupling (string theory) perturbative expansions, and confirms the need of a solid explanation of their origin.

\section{Analysis and results}

\begin{description}

\item \emph{a) Closed formulas for the anomalous dimension}

At one-loop, the gluonic sector is described by the $XXX_{-3/2}$ closed spin chain, and the anomalous dimension is known as an exact solution of the Baxter equation. At higher orders, we solve perturbatively the long range Bethe equations whose compact form is
\be\label{Bethe}
\left(\frac{u_j+\frac{i}{2}V_{k_j}}{u_j-\frac{i}{2}V_{k_j}}\right)^L =
\mathop{\prod_{\ell=1}^K}_{\ell\neq j} \frac{u_j-u_\ell+\frac{i}{2}M_{k_j,k_\ell}}{u_j-u_\ell-\frac{i}{2}M_{k_j,k_\ell}}.
\ee
where $M_{k\ell}$ is the Cartan matrix of the algebra and $V_k$ 
are the Dynkin labels of the spin representation carried by each site of the chain~\footnote{Together with the equations (\ref{Bethe}), one has to consider the additional costraint given from the cyclicity of the spin chain.}. The excitation numbers $K_i$ of the Bethe roots $u_i$ can be computed~\cite{Beisert:2003yb} from  the quantum numbers of the superconformal state associated with the twist-3 gluonic operator of interest here. In order to identify the correct superconformal primary describing this sector, one can exploit the superconformal properties of the (maximally symmetric) tensorial product of three \emph{singletons}~\cite{Beisert:2004di}. This has been done in Ref.~\cite{Beccaria:2007pb}, where the Dynkin diagram associated to the Cartan matrix in (\ref{Bethe}) for the case of interest here has been found to be
\be
\begin{minipage}{260pt}
\setlength{\unitlength}{1pt}
\small\thicklines
\begin{picture}(260,55)(-10,-30)
%
\put(-32,0){\line(1,0){22}}  
\put(  0,00){\circle{15}}
\put( -5,-5){\line(1, 1){10}}  
\put( -5, 5){\line(1,-1){10}}  
\dottedline{3}(8,0)(32,0)    
\put( 40,00){\circle{15}}     
\dottedline{3}(48,0)(72,0)   
\put( 80,00){\circle{15}}
\put( 75,-5){\line(1, 1){10}}  
\put( 75, 5){\line(1,-1){10}}  
\put( 80,-15){\makebox(0,0)[t]{$N+3$}}  
\put( 87,00){\line(1,0){26}} 
\put(120,00){\circle{15}}
\put(120,15){\makebox(0,0)[b]{$+1$}} 
\put(120,-15){\makebox(0,0)[t]{$N+4$}} 
\put(127,00){\line(1,0){26}} 
\put(160,00){\circle{15}}
\put(155,-5){\line(1, 1){10}}  
\put(155, 5){\line(1,-1){10}}  
\put(160,-15){\makebox(0,0)[t]{$N+2$}} 
\dottedline{3}(168,0)(192,0) 
\put(200,00){\circle{15}}
\put(200,-15){\makebox(0,0)[t]{$1$}} 
\dottedline{3}(208,0)(232,0) 
\put(240,00){\circle{15}}
\put(235,-5){\line(1, 1){10}} 
\put(235, 5){\line(1,-1){10}} 
\put(250,0){\line(1,0){20}} 
\end{picture}
\end{minipage}
\ee
The number on top of the diagram indicates the spin representation,  the numbers below are the root excitation numbers of the superconformal primary. 
Using the one-loop solution as an input~\footnote{See Ref.~\cite{Beccaria:2007pb} for an explanation of the necessary (backtraced) dualization of the Bethe roots.},  one can  expand the Bethe equations  in the coupling constant g order by order in perturbation theory. 
The equations for the quantum corrections to the one-loop roots are linear, and thus numerically solvable with high precision. The resulting anomalous dimension has rational coefficients in its loop expansions, that can be easily and unambiguously identified according to the methods discussed in~ Refs.~\cite{Beccaria:2007cn,Kotikov:2007cy}. In order to find a suitable closed analytical formula for the first loops, one can assume for the anomalous dimension a generalized form of the maximum transcendentality principle. Inspired by  the one-loop result~\cite{Beisert:2004di}, in which not all the terms present a constant degree of transcendentality~\footnote{An alternative point of view is to adopt the maximum transcendentality principle in a related, non-canonical, basis of harmonic sums.} and by  similar QCD calculations~\cite{Mertig:1995ny},  the following Ansatz can be made which generalizes the one-loop result
\be
\label{Ansatz}
\gamma_n = \sum_{\tau=0}^{2\,n-1}\gamma_n^{(\tau)}, ~~~~~
\gamma_n^{(\tau)} = \sum_{k+\ell = \tau}\frac{{\cal H}_{\tau,\ell}(n)}{(n+1)^k}, \qquad n = \frac{N}{2}+1,\nonumber
\ee
where ${\cal H}_{\tau,\ell}(n)$ is a combination of harmonic sums with homogeneous fixed transcendentality $\ell$. The terms with $k=0$ have maximum  transcendentality, all the others have subleading  transcendentality.
The three loop result has been found in Ref.~\cite{Beccaria:2007pb}. In Ref.~\cite{BF}, we have computed a long list of values for the four-loop anomalous dimension  $\gamma_4(n)$ as exact rational numbers obtained from the perturbative expansion of the long-range Bethe equations. We have matched them against the general Ansatz (\ref{Ansatz}). A very large number of possible terms appear with unknown
coefficients. To reduce them, we have imposed some structural properties emerged from the analysis of the three loop result (see details in Ref.~\cite{BF}).
Deferring the reader to Ref.~\cite{BF} for the complete result, we report here, using the notation of (\ref{Ansatz}), the term with maximal trancendentality~\footnote{A further transcendentality $7$ term comes from the contribution of the so-called dressing factor in the Bethe equations and consists of a combination of harmonic sums of transcendentality $4$  multiplied by the characteristic $\zeta_3$ contribution.}
\begin{eqnarray}
{\cal H}_{7,7} &=& 
\frac{S_7}{2}+7 \,S_{1,6}+15 \,S_{2,5}-5 \,S_{3,4}-29 \,S_{4,3}-21 \,S_{5,2}-5
   \,S_{6,1}-40 \,S_{1,1,5} + \nonumber \\
&&\!\!\!\!\!\!\!\!\!\!\!\!-32 \,S_{1,2,4}+24 \,S_{1,3,3}+ + 32 \,S_{1,4,2}-32 \,S_{2,1,4}+20
   \,S_{2,2,3}+40 \,S_{2,3,2} +4 \,S_{2,4,1} \nonumber \\
&&\!\!\!\!\!\!\!\!\!\!\!\!+24 \,S_{3,1,3}+44 \,S_{3,2,2}+24
   \,S_{3,3,1} + 36 \,S_{4,1,2}+36 \,S_{4,2,1}  +24 \,S_{5,1,1}+80 \,S_{1,1,1,4}+\nonumber \\
&&\!\!\!\!\!\!\!\!\!\!\!\!-16
   \,S_{1,1,3,2}+32 \,S_{1,1,4,1}-24 \,S_{1,2,2,2}+16 \,S_{1,2,3,1} -24 \,S_{1,3,1,2} -24
   \,S_{1,3,2,1}+\nonumber \\
&&\!\!\!\!\!\!\!\!\!\!\!\!-24 \,S_{1,4,1,1}-24 \,S_{2,1,2,2}+16 \,S_{2,1,3,1}-24 \,S_{2,2,1,2}-24
   \,S_{2,2,2,1}  -24 \,S_{2,3,1,1}+  \nonumber \\
&&\!\!\!\!\!\!\!\!\!\!\!\!-24 \,S_{3,1,1,2}-24 \,S_{3,1,2,1}-24 \,S_{3,2,1,1}-24
   \,S_{4,1,1,1}-64 \,S_{1,1,1,3,1}
\end{eqnarray}
   where $S_a\equiv S_a(n)$ with $n = \frac{N}{2}+1$ are the nested harmonic sums defined by
\begin{equation}
S_a(N) = \sum_{n=1}^N\frac{1}{n^a},~~~~~~~~~~~~~~~~~
S_{a, \mathbf{b}}(N) = \sum_{n=1}^N\frac{1}{n^a}\, S_{\mathbf b}(n).
\end{equation}

\item \emph{b) Reciprocity respecting formulas}

Proving  reciprocity for the gluonic operators amounts to first deriving the $\P$ function via an inversion of the relation (\ref{nonlinear}), which in terms of  the perturbative expansions
${\cal P} = \sum_{k=1}^\infty {\cal P}_k\,g^{2\,k}$ and $\gamma = \sum_{k=1}^\infty \gamma_k\,g^{2\,k}$ 
 reads eventually as follows
\begin{eqnarray}\label{pertP}
{\cal P}_1 = \gamma_1, ~~~~~~~{\cal P}_2 = \gamma_2-\frac{1}{8}\,(\gamma_1^2)',....
\end{eqnarray}
Secondly, one has to check the parity invariance (\ref{parity}) with respect of the quadratic Casimir which in this case is 
\be
J^2 = N^2 + 8\,N + \frac{63}{4} = 4\,n(n+2)+\frac{15}{4}.
\ee
The constant above is irrelevant to the proof and one can define an effective Casimir
$
J^2_{\rm eff}=n\,(n+2).
$
Remarkably, a complete proof of reciprocity at four loops can be given in closed form. To this aim, the following observations (for their proofs see Ref.~\cite{BF}) have been useful 
\begin{itemize}
\item
{\bf Theorem 2.1.} {\em
Let $f(n)$ be reciprocal with respect to $J^2 = n(n+1)$. Then, the combination 
$\widetilde{f}(n) = f(n) + f(n+1)$ is reciprocal with respect to $J^2_{\rm eff}$.
}

Considering then the   \emph{linear map} defined on linear combinations of simple $S$ sums by 
$
\Phi_a (S_{b, \mathbf{c}}) = S_{a, b, \mathbf{c}}-\frac{1}{2}\, S_{a+b, \mathbf{c}}
$
and defining
\be\nonumber
I_a = S_a,~~~~~~~~~~
I_{a_1, a_2, \dots, a_n} = \Phi_{a_1}(\Phi_{a_2}(\cdots\, \Phi_{a_{n-1}}(S_{a_n})\cdots )~~~~{\rm one }~~{\rm finds}
\ee
\item
{\bf Theorem 2.2.} {\em
The combinations $I_{a_1, \dots, a_n}$ with odd $a_1, \dots, a_n$ have a large $N$ RR expansion
\be
I_{a_1, \dots, a_n} = \sum_{\ell=0}^\infty \frac{P_\ell(\log\,J^2)}{J^{2\ell}},
\ee
where $J^2 = N(N+1)$ and $P_\ell$ is a polynomial.
}
\end{itemize}

These observations can be exploited writing the function $\P$ in (\ref{pertP}) in a manifestly RR form. For example, at three loops
\ba\label{P3}
 \!\!\! \!\!\! \!\!\!{\cal P}_3 &=&
\frac{\,\widetilde{I}_3}{2 (n+1)^2}+\frac{3 \,\widetilde{I}_5}{2}-4 \,\widetilde{I}_{1,1,3}
+\frac{2}{(n+1)^4}-4 \,\widetilde{I}_{1,3}
+\frac{\pi ^2 \,\widetilde{I}_3}{6} -2 \,\widetilde{I}_3+4 \,\widetilde{I}_{1,1} \zeta _3+ \nonumber\\
&& 
-\frac{\zeta _3}{(n+1)^2}-\frac{4}{(n+1)^2}
+4 \zeta _3 \,\widetilde{I}_1+\frac{4 \pi ^4 \,\widetilde{I}_1}{45} +4 \zeta _3+\frac{8 \pi ^4}{45}+\frac{4 \pi ^2}{3}+32 .
\ea
where $\widetilde{I}_\mathbf{a}\equiv \widetilde{I}_\mathbf{a}(n)=\widetilde{I}_\mathbf{a}(n)+\widetilde{I}_\mathbf{a}(n+1)$.
As explained in details in Ref.~\cite{BF}, reciprocity is evident because formula (\ref{P3}) is a combinations of invariants (as from Theorem~2.2) $\widetilde{I}_\mathbf{a}$~\footnote{Invariants  $\widetilde{I}_\mathbf{a}$ have to appear with odd indices~\cite{BF}.}   and factors $(n+1)^{-p}$ with even $p$~\footnote{The constraint on $p$ is due to the relation $n+1 = \sqrt{n\,(n+2)+1}$.}. In this case, the expression is automatically reciprocity respecting with respect to $n(n+2)$. Analogous manifestly reciprocity invariant expressions for  $\P$ up to four loops are collected in Ref.~\cite{BF}.
\end{description}

 

\end{document}